\documentclass[Physsubmission, Phys, a4paper]{SciPost}

\hypersetup{
    colorlinks,
    linkcolor={red!50!black},
    citecolor={blue!50!black},
    urlcolor={blue!80!black}
}

\usepackage[bitstream-charter]{mathdesign}
\DeclareSymbolFont{usualmathcal}{OMS}{cmsy}{m}{n}
\DeclareSymbolFontAlphabet{\mathcal}{usualmathcal}

\begin{document}

\begin{center}{\Large \textbf{
Sub-TeV hadronic interaction model differences and their impact on air showers
}}\end{center}

\begin{center}
M. Schmelling\textsuperscript{1$\star$}, 
\'A. Pastor-Guti\'errez\textsuperscript{1},
H. Schorlemmer\textsuperscript{2,3,1}, 
R.D. Parsons\textsuperscript{4,1}
\end{center}

\begin{center}
{\bf 1} Max-Planck-Institut f\"ur Kernphysik, Heidelberg, Germany
\\
{\bf 2} IMAPP\,\!, Radboud University Nijmegen, Nijmegen, The Netherlands
\\
{\bf 3} Nationaal Instituut voor Kernfysica en Hoge Energie Fysica (NIKHEF), Science Park, Amsterdam, The Netherlands
\\
{\bf 4} Institut f\"ur Physik, Humboldt-Universit\"at zu Berlin, Berlin, Germany
\\
* michael.schmelling@mpi-hd.mpg.de
\end{center}

\begin{center}
\today
\end{center}


\definecolor{palegray}{gray}{0.95}
\begin{center}
\colorbox{palegray}{
  \begin{tabular}{rr}
  \begin{minipage}{0.1\textwidth}  
    \includegraphics[width=23mm]{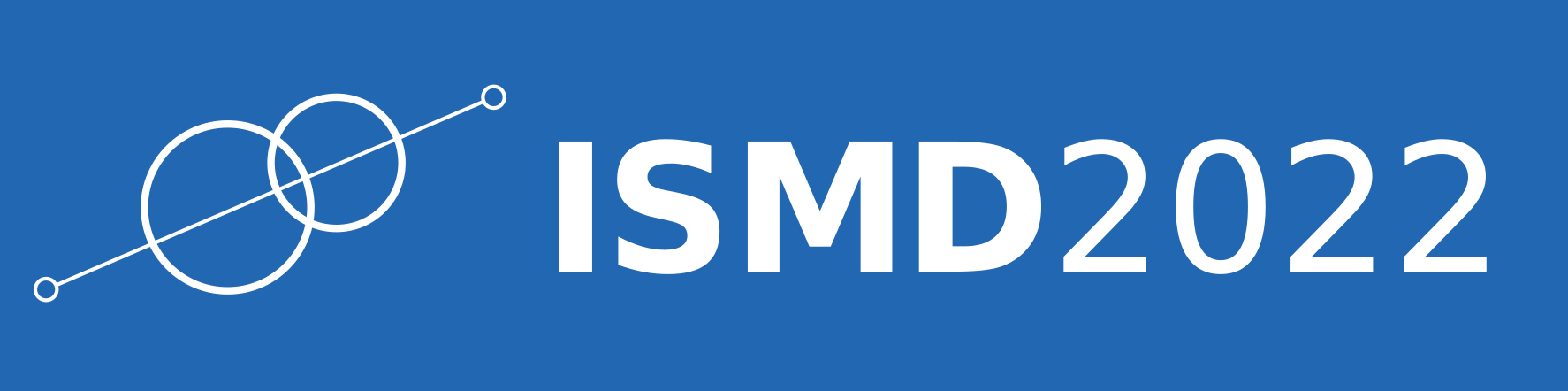} 
  \end{minipage}
  &
  \begin{minipage}{0.8\textwidth}  
    \begin{center}
    {\it 51st International Symposium on Multiparticle Dynamics (ISMD2022)}\\ 
    {\it Pitlochry, Scottish Highlands, 1-5 August 2022} \\
    \doi{10.21468/SciPostPhysProc.?}\\
    \end{center}
  \end{minipage}
\end{tabular}
}
\end{center}

\section*{Abstract}
{\bf
In the sub-TeV regime, the most widely used hadronic interaction models 
disagree significantly in their predictions of particle spectra from cosmic 
ray induced air showers. We investigate the nature and impact of model uncertainties,
focussing on air shower primaries with energies around the transition between 
high and low energy hadronic interaction models, where the dissimilarities are
largest and which constitute the bulk of the interactions in air showers.}

\section{Introduction}
\label{sec:intro}
The description of air showers created by  high energy cosmic ray primaries 
hitting the atmosphere requires the modelling of hadronic interactions 
between elementary particles and air nuclei over many orders of magnitude 
in energy. The need to understand hadronic physics over such a large 
energy range is highlighted for example by the observation of an 
unexpectly large muon flux in high energy cosmic-ray interactions 
\cite{Dembinski2019, Albrecht2022}, or in estimates of the physics 
reach of future astroparticle physics experiments \cite{Ohishi2021}. 
Model comparisons focussing on primaries with energies in the range 
from 100 GeV to 100 TeV are discussed in \cite{Parsons2019}, and a study 
of how model predictions compare to experimental data recorded at the LHC 
can be found e.g.~in \cite{Anchordoqui2020}. However, the fact that 
in extensive air showers the bulk of the particle production happens 
late in the shower evolution puts emphasis also the low energy region 
below 100 GeV. This has been studied in \cite{Pastor2021}, some key 
results of which are summarised below. 

\begin{figure}[t]
\centering
\includegraphics[width=0.75\textwidth]{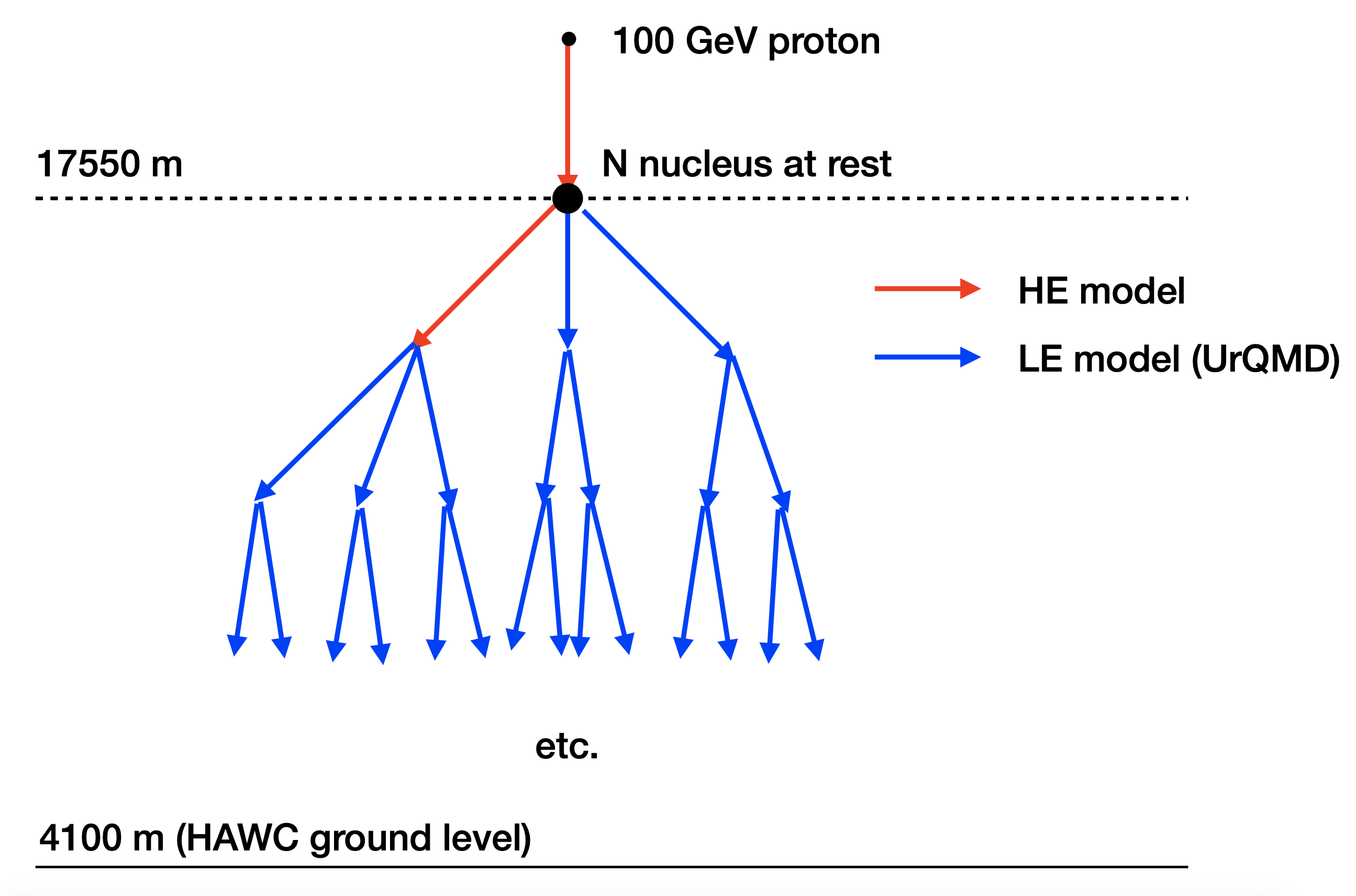}
\caption{Sketch of the scenario considered in this study.}           
\label{fig:cascade}
\end{figure}

The scenario considered is sketched in fig.\;\ref{fig:cascade} and modelled 
by the CORSIKA v7.64 \cite{CORSIKA} air shower simulation software package. 
A primary proton with a total lab energy of 100 GeV and zero zenith angle
interacts with a nitrogen nucleus at an altitude of 17550\,m. The observation 
level for the final state is at 4100\,m. Fixing the height of the first 
interaction removes geometric effects caused by varying ground distances and 
puts the focus on differences between the physics modelling. Here EPOS-LHC 
\cite{EPOS-LHC}, QGSJetII-04 \cite{QGSJetII-04}, SIBYLL~2.3c \cite{SIBYLL2.3c} 
and UrQMD \cite{UrQMD} are considered. UrQMD is designed for lab energies from 
less than 100\,MeV up to O(200)\,GeV, the  other models can be applied for lab 
energies above O(40)\,GeV. A full shower simulation starts with a high-energy 
(HE) model for the initial part of the shower development and switches to a 
low-energy (LE) model at a transition energy, which in CORSIKA is set to 
80 GeV. With a 100 GeV primary all four models are suitable for the initial 
interaction, the evolution below the transition energy is modelled by UrQMD.

\section{Event classification}
Ground level observables are affected by the interplay of what happens
in the first, highest energy, interaction and the subsequent lower
energy processes. Here the first interaction will be characterised by 
the inelasticity $\kappa$ of the event and the type of the leading particle, 
i.e. the secondary with the highest energy. The inelasticity is defined as
\begin{equation}
   \kappa = 1 - \frac{E_{\rm LP}}{E_{\rm FI}} \approx 1 - x_F^{\rm LP} \;,
\end{equation}
where $E_{\rm LP}$ is the energy of the leading particle and $E_{\rm FI}$ the 
total energy of all final state particles. This definition ensures that 
$\kappa$ is in the range $[0,1]$ and is insensitive to small violations
of energy momentum conservation that are observed in all models.  The 
inelasticity is related to Feyman's scaling variable $x_F$ and provides a 
qualitative measure for the amount of energy that goes into the 
production of  new particles.

\begin{figure}[tb]
\centering
\begin{minipage}{0.575\textwidth}
\centering
\includegraphics[width=\textwidth]{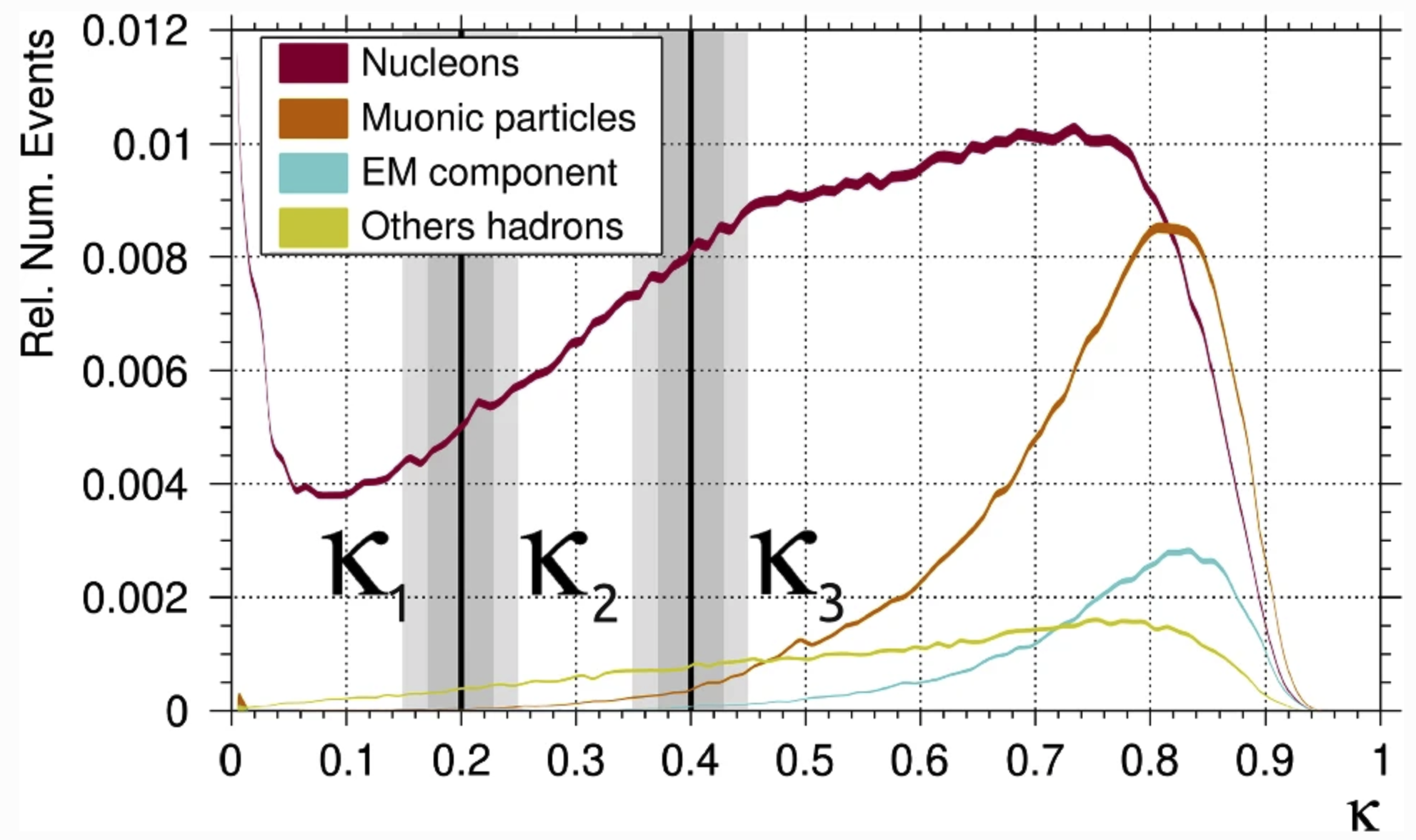}
\end{minipage}
\begin{minipage}{0.375\textwidth}
\begin{flushright}
\begin{tabular}{ll}
 Nucleons:       & $p, n$ \\
 Muonic family:  & $\mu,\pi,K$ \\
 EM component:   & $\gamma, e$        \\ 
 Others:         & $\Lambda,\Sigma,\Xi,\Omega\ldots$
\end{tabular}
\end{flushright}
\end{minipage}
\caption{Inelasticity distributions for events with different types of leading particles 
 in EPOS-LHC, and definition of the particle families used to classify events. 
 Antiparticles and different charge states are implied. The coloured bands in the
 plot visualise the statistical uncertainties of the simulation, the shaded regions 
 indicate the systematic uncertainties in the definition of $\kappa$ due to 
 event-by-event fluctuations in the amount of energy violation discussed in the 
 appendix.}
\label{fig:kappa-epos}
\end{figure}

For the leading particle we differentiate between nucleons, the muonic family, 
which contains particles that either directly or via decays contribute to the 
ground level muon flux, the EM component, which leads to electromagnetic showers 
and others, which are less important for the ground level observables. 
Figure\;\ref{fig:kappa-epos} shows the inelasticity distribution in EPOS-LHC
for events with different types of leading particles. For small inelasticities 
the subsequent shower evolution is driven by leading nucleons, particles from 
the muonic family become important at large inelasticities. In the following 
we consider three regions: the elastic and diffractive region $\kappa_1\in[0,0.2]$,  
the transition region $\kappa_2\in[0.2,0.4]$, and the highly inelastic regime 
$\kappa_3\in[0.4,1]$. 

\section{Ground level observables}
\subsection{Lateral distributions of muons and electromagnetic energy flow}
\begin{figure}[tb]
\centering
\includegraphics[width=0.9\textwidth]{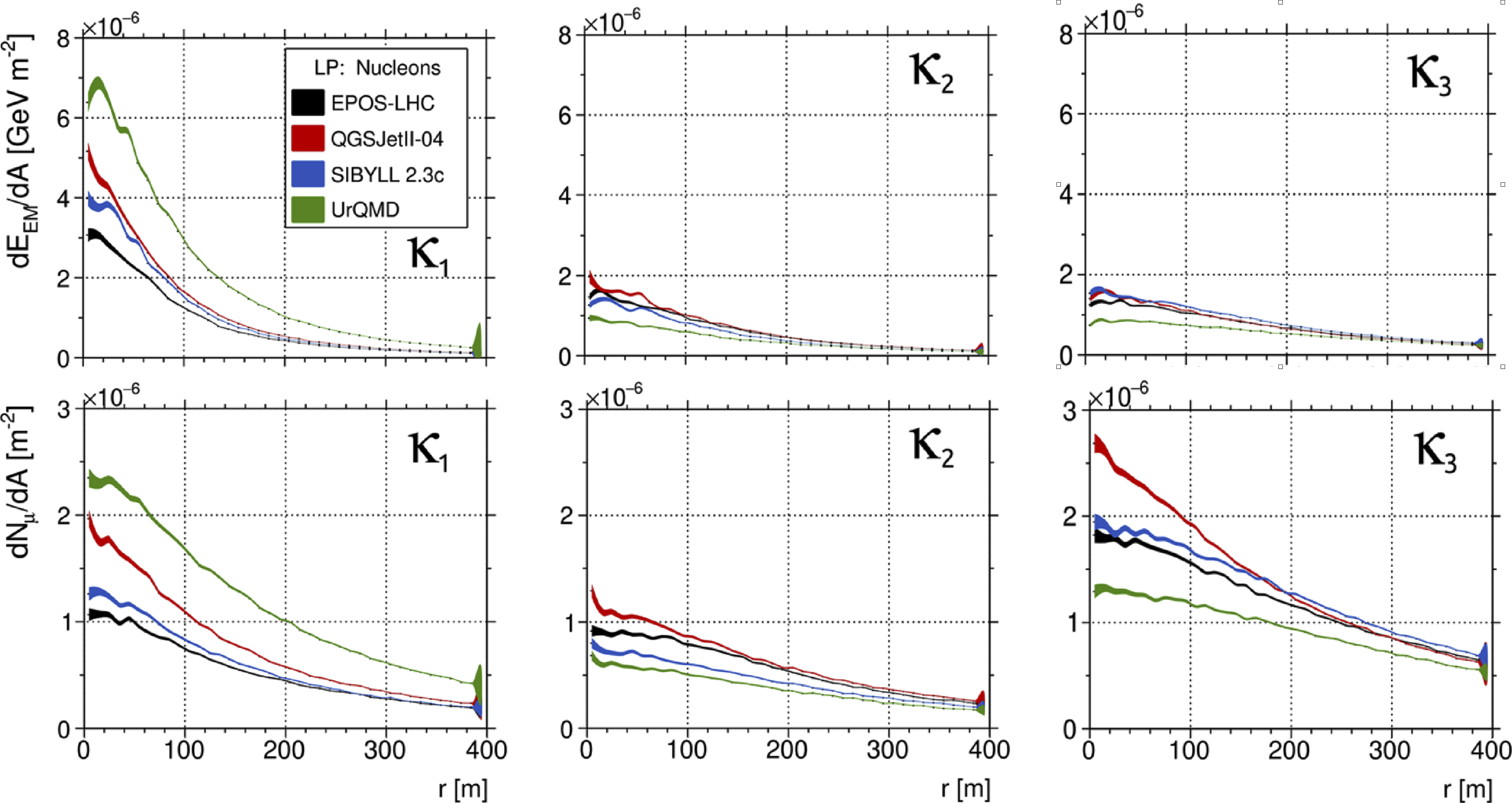}
\caption{Lateral distributions electromagnetic energy flow (top row) and
        of muon numbers (bottom row) for different inelasticity regions 
        and different HE interaction models.}
\label{fig:ldf}
\end{figure}

Figure\;\ref{fig:ldf} shows the lateral distributions of muons and the
electromagnetic energy flow at ground level when the leading particle 
of the first interaction is a nucleon. As a function of the distance 
to the shower axis, the muon flux is shown in units of muons per square 
meter, for the electromagnetic component the energy deposit is given 
in GeV per square meter.

The electromagnetic energy flow is dominated by events from the diffractive 
region since in events with small inelasticity the bulk of the particle 
production happens deeper in the atmosphere, with the consequence that more 
of the electromagnetically interacting particles reach the ground. For the 
same reason also a sizeable fraction of the muon flux comes from the diffractive 
$\kappa_1$ region. In addition there is a large contribution from highly 
inelastic events, where many particles are created that decay into muons.
The model with the largest number of muons close to the shower centre is 
QGSJetII-04. The excess is even more pronounced for events with a leading 
particle from the muonic family \cite{Pastor2021}.

\subsection{Muon flux at ground level}
\begin{figure}[t]
\centering
\raisebox{1.5mm}{\includegraphics[width=0.48\textwidth]{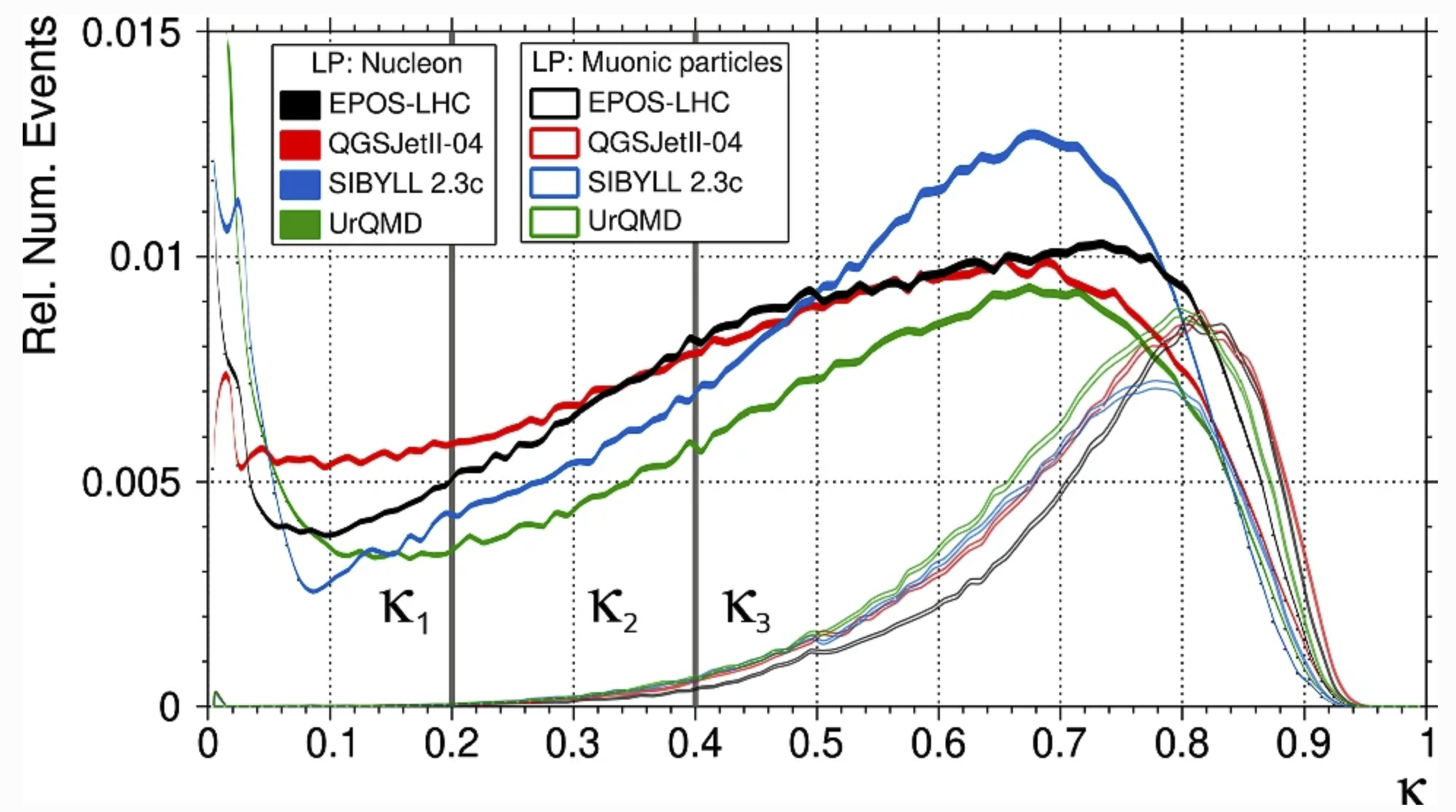}}
\hspace{2mm}
\includegraphics[width=0.47\textwidth]{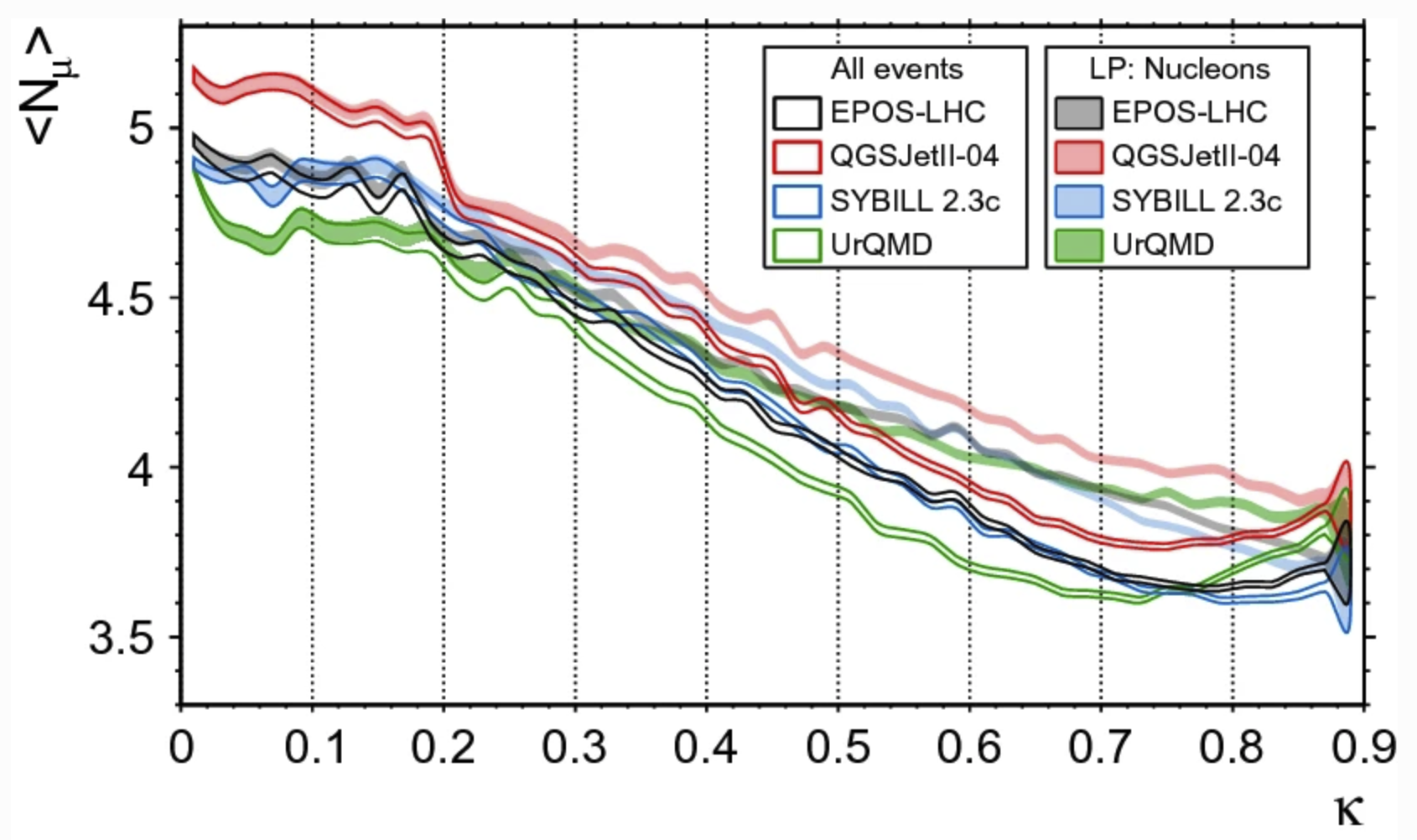}
\caption{Inelasticity distributions (left) of the first interaction
         for events where the leading particle if a nucleon or belongs
         to the muonic family, and (right) average number of ground-level 
         muons per event as a function of the inelasticity $\kappa$. The 
         filled bands are for events with leading nucleons in the first 
         interaction, the open ones are for all events.} 
\label{fig:muons}
\end{figure}

The overall numbers of muons at ground level are more similar since the area 
at small distances from the shower centre is only small fraction of the total. 
The average number of muons per event as a function of the inelasticity is 
shown in fig.\;\ref{fig:muons}, together with the $\kappa$ distributions of 
those events that dominate the muon flux at ground level, i.e. those with a 
leading nucleon or a leading particle from the muonic family. In the muon 
numbers the HE models show a discontinuity at $\kappa\approx0.2$. Below that
value the first two interactions, above only the first interaction is modelled 
by one of EPOS-LHC, QGSJetII-04 or SIBYLL~2.3c. The spread of the curves 
for $\kappa>0.2$ thus shows the importance of the first interaction for 
ground level observables, the jump at $\kappa\approx 0.2$ reflects the 
differences of the physics modelling between the HE models and UrQMD at the
transition energy. The difference is largest for QGSJetII-04. Smaller but
still significant jumps are observed for EPOS-LHC and SIBYLL~2.3c.
  
\section{Conclusions}
Studies of hadronic interaction models in the transition region between
the HE and LE regime reveal significant differences in the prediction
of ground level observables. Those can be traced to differences in the 
final states of the first interaction generated by the HE models and
a discontinuity in the physics modelling when switching from a HE model
to the LE model. Improvements of the models are expected from comparing 
their predictions to existing and upcoming data on inclusive particle 
production cross-sections from accelerator experiments. Here nucleon-nucleon 
centre-of-mass energies $\sqrt{s_{NN}}$ of $O(10)$\,GeV are probed by 
e.g.~NA61/SHINE at the CERN-SPS, of $O(100)$\,GeV by LHCb fixed target 
data and $O(10)$\,TeV by proton-proton and proton-lead collisions recorded 
by the ALICE, ATLAS, CMS and LHCb experiments. Key data 
\cite{Dembinski2020} for the understanding of cosmic-ray induced 
air showers is also expected from measurements of proton-oxygen 
and oxygen-oxygen collisions that are scheduled for Run 3 of the LHC.


\begin{appendix}
\section{Energy-momentum violation in air shower models}
\begin{figure}[htb]
\centering
\includegraphics[width=0.95\textwidth]{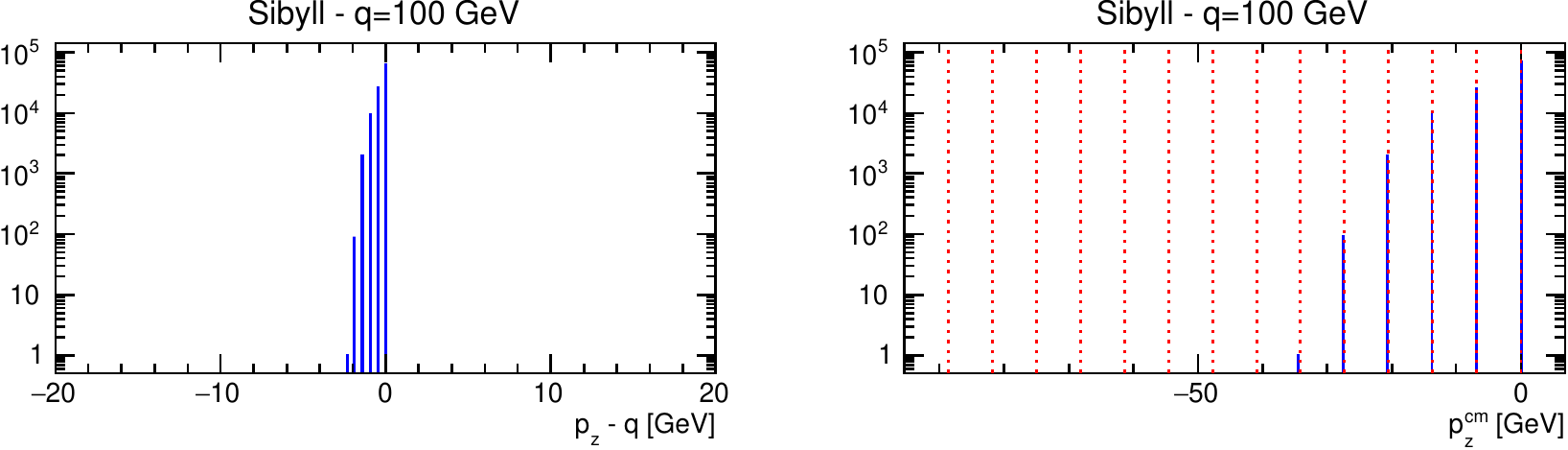}
\includegraphics[width=0.95\textwidth]{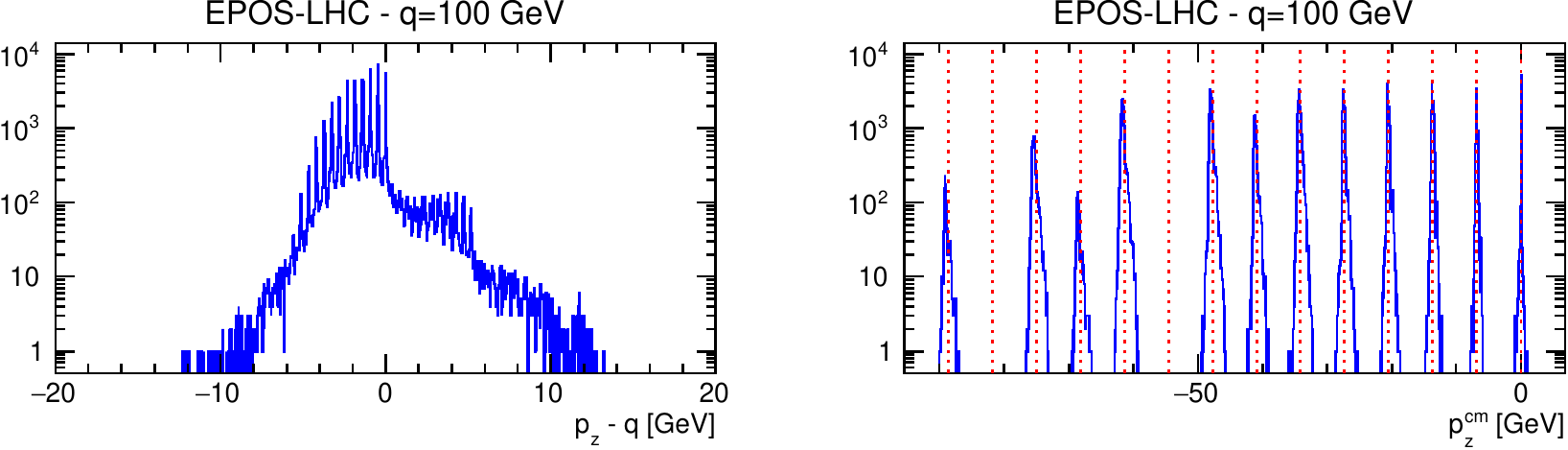}
\caption{Violation of momentum conservation in SIBYLL~2.3C and EPOS-LHC. The left hand
         plots show the effect in the lab system, the right hand plots in the 
         nucleon-nucleon centre-of-mass system. The red dashed lines indicate
         the expected values when $N=1,2,\ldots$ target nucleons are involved in
         the interaction.}
\label{fig:eviol}
\end{figure}

Using the CRMC interface \cite{CRMC}, the hadronic interaction models can 
be run standalone in order to compare to accelerator data or for simple checks of
the kinematics. Taking for example an incoming proton with a lab momentum
 $q=100$\,GeV/$c$ colliding with a nucleus at rest, momentum conservation 
requires that the momentum sum of all final state particles is $100$\,GeV/$c$. 
As shown in fig.\,\ref{fig:eviol} there are deviations. 

In SIBYLL-2.3c the effects are small, with slight offsets that are related to 
the number of target nucleons involved in the interaction. This can be seen by 
boosting the event to the nucleon-nucleon centre-of-mass, where the actual 
event generation happens. For an interaction with $N>1$ target nucleons and a
boost that goes to the centre-of-mass for $N=1$, the boost generates a momentum 
excess of $1-N$ times the centre-of-mass momentum of the incoming proton. 
For SIBYLL-2.3c one finds delta-functions that are marginally displaced from 
the expected values and correspond to the small shifts seen in the lab system.

For EPOS-LHC the centre-of-mass distributions show a sizeable smearing around 
the expected values, which correspond to violations of momentum conservation
of up to $10$\,GeV/$c$ in the lab system. The large deviations can be avoided 
by improving the numerical precision of the model, changing in line 17 of the 
function {\tt epos-uti.f} the value {\tt 0.5} to e.g.{\tt 0.005} \cite{Tanguy}:\\[1mm]
\centerline{\tt if(iLHC.eq.1) errlim=max(0.00005,0.5/engy)} 

\vspace*{1mm}
The boost to the nucleon-nucleon centre-of-mass allows one also to determine 
the number of interacting target nucleons. Here, in SIBYLL-2.3c only up to 
5 target nucleons interact whereas in EPOS-LHC the full range from 1--14 
except for N=9 and N=13  is covered. 
  
\begin{figure}[htb]
\centering
\includegraphics[width=0.75\textwidth]{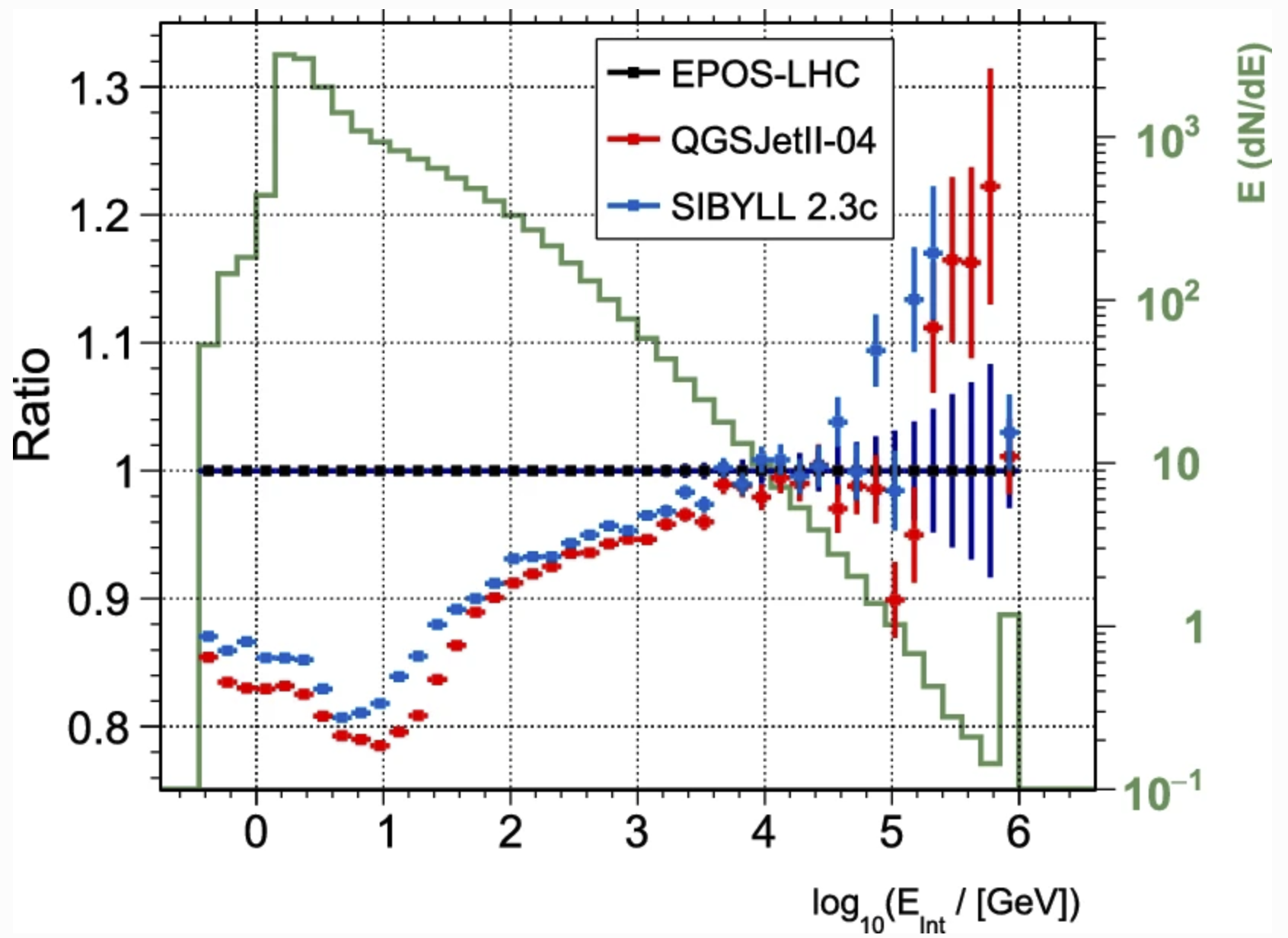}
\caption{Average number of interactions as a function of the particle
         energy in air showers initiated by 1 PeV protons hitting the 
         atmosphere. The histogram (green, right hand ordinate) is 
         the prediction by EPOS-LHC, the points (left hand ordinate) 
         show the ratios to EPOS-LHC for SIBYLL~2.3c and QGSJetII-04.} 
\label{fig:freq}
\end{figure}

The importance of correctly modelling sub-TeV hadronic interactions for 
air showers is underlined by fig.\,\ref{fig:freq}, which, for 1 PeV primaries, 
shows the average number of interactions in the shower as a function of 
the particle energy. The number of interactions rises exponentially as the 
particle energy drops, such that the bulk of the particle production occurs 
at energies at or below the transition energy. The plot also shows that the 
number of sub-TeV interactions in SIBYLL~2.3c and QGSJetII-04 is up to
20\% lower than for EPOS-LHC.

\end{appendix}



\nolinenumbers

\begin{thebibliography}{99}
\bibitem{Dembinski2019}
\mbox{H.~P.~Dembinski et al.},
{\it Report on Tests and Measurements of Hadronic Interaction Properties with Air Showers},
EPJ Web of Conferences {\bf 210}, 02004 (2019),\\
\url{https://doi.org/10.1051/epjconf/201921002004}.

\bibitem{Albrecht2022}
\mbox{J.~Albrecht et al.},
{\it The Muon Puzzle in cosmic-ray induced air showers and its 
connection to the Large Hadron Collider},
Astrophys.Space Sci. {\bf 367} (3) 27 (2022),
\url{https://doi.org/10.1007/s10509-022-04054-5}.

\bibitem{Ohishi2021}
\mbox{M.~Ohishi et al.}, 
{\it Effect of the uncertainty in the hadronic interaction models
on the estimation of the sensitivity of the Cherenkov Telescope Array}, 
J. Phys. G: Nucl. Part. Phys. {\bf 48} 075201 (2021),
\url{https://doi.org/10.1088/1361-6471/abfce0}.

\bibitem{Parsons2019}
\mbox{R.D.~Parsons et al.}, 
{\it Systematic differences due to high energy hadronic interaction 
models in air shower simulations in the 100 GeV‚Äì100 TeV range}, 
Phys. Rev. D {\bf 100}, 023010 (2019), 
\url{https://doi.org/10.1103/PhysRevD.100.023010}.

\bibitem{Anchordoqui2020}
\mbox{L.~A.~Anchordoqui et al.},
{\it Through the Looking-Glass with ALICE into the Quark-Gluon Plasma: 
A New Test for Hadronic Interaction Models Used in Air Shower Simulations},
Phys. Lett. B {\bf 810}, 135837 (2020),
\url{https://doi.org/10.1016/j.physletb.2020.135837}.

\bibitem{Pastor2021}
\mbox{\'A.~Pastor-Guti\'errez et al.}, 
{\it Sub-TeV hadronic interaction model differences and their impact on air showers}, 
Eur.Phys.J. C {\bf 81} 369 (2021),\\
\url{https://doi.org/10.1140/epjc/s10052-021-09160-2}.

\bibitem{CORSIKA}
\mbox{D.~Heck et al.},
{\it CORSIKA : A Monte Carlo Code to Simulate Extensive Air Showers}, 
For\-schungs\-zentrum Karlsruhe, Report FZK 6019 (1998),\\
\url{https://web.iap.kit.edu/corsika/physics\_description/corsika\_phys.pdf}.

\bibitem{EPOS-LHC}
\mbox{T.~Pierog et al.}, 
{\it EPOS-LHC: test of collective hadronization with 
data measured at the CERN large hadron collider} Phys. Rev. C {\bf 92}, 034906 (2015),\\
\url{https://doi.org/10.1103/PhysRevC.92.034906}.

\bibitem{QGSJetII-04}
\mbox{S.~Ostapchenko}, 
{\it Monte Carlo treatment of hadronic interactions in
 enhanced Pomeron scheme: QGSJET-II model}, 
 Phys. Rev. D {\bf 83}, 014018 (2011),\\
\url{https://doi.org/10.1103/PhysRevD.83.014018}.

\bibitem{SIBYLL2.3c}
\mbox{F.~Riehn et al.}, 
{\it The hadronic interaction model Sibyll 2.3c and Feynman scaling}, 
35th International Cosmic Ray Conference, PoS(ICRC2017), {\bf 301} (2017),\\
\url{https://doi.org/10.22323/1.301.0301}.

\bibitem{UrQMD}
\mbox{S.~Bass et al.}, 
{\it Microscopic models for ultrarelativistic heavy ion collisions}, 
Prog. Part. Nucl. Phys. {\bf 41} 255 (1998), 
\url{https://doi.org/10.1016/S0146-6410(98)00058-1}. 

\bibitem{Dembinski2020}
\mbox{H.~P.~Dembinski et al.},
{\it Future Proton-Oxygen Beam Collisions at the LHC for Air Shower Physics},
PoS ICRC2019, 235 (2020),
\url{https://doi.org/10.22323/1.358.0235}.

\bibitem{CRMC}
\mbox{R.~Ulrich et al.}, {\it Cosmic Ray Monte Carlo Package, CRMC},\\
\url{https://doi.org/10.5281/zenodo.4558706}. 

\bibitem{Tanguy}
T.~Pierog, private communication.
\end{thebibliography}
\end{document}